\documentclass[useAMS,usenatbib,twocolumn]{mn2e}
\usepackage{natbib}
\usepackage{graphicx}
\usepackage{amssymb,latexsym,amsopn}
{\tiny }\usepackage{epsfig}
\usepackage[T1]{fontenc}

\usepackage{color}

\usepackage{ulem}

\title[Supramassive neutron star collapse times]
{The birth of black holes: neutron star collapse times, gamma-ray bursts and fast radio bursts}
\author[Ravi \& Lasky]
  {Vikram Ravi$^{1,2}$\thanks{v.vikram.ravi@gmail.com} and Paul D. Lasky$^{1}$\thanks{paul.lasky@unimelb.edu.au}\\
  	$^{1}$School of Physics, University of Melbourne, Parkville, VIC 3010, Australia\\
  	$^{2}$CSIRO Astronomy and Space Science, Australia Telescope National Facility, P.O. Box 76, Epping, NSW 1710, Australia 
  }

\date{\today}

\pagerange{1--8} \pubyear{2014}

\def\LaTeX{L\kern-.36em\raise.3ex\hbox{a}\kern-.15em
    T\kern-.1667em\lower.7ex\hbox{E}\kern-.125emX}

\begin{document}

\label{firstpage}

\maketitle

\begin{abstract}

Recent observations of short gamma-ray bursts (SGRBs) suggest that binary neutron star (NS) mergers can create highly magnetised, millisecond NSs.  Sharp cut-offs in $X$-ray afterglow plateaus of some SGRBs hint at the gravitational collapse of these remnant NSs to black holes. 
The collapse of such `supramassive' NSs also describes the blitzar model, a leading candidate for the progenitors of fast radio bursts (FRBs). The observation of an FRB associated with an SGRB would provide compelling evidence for the blitzar model and the binary NS merger scenario of SGRBs, and lead to interesting constraints on the NS equation of state. We predict the collapse times of supramassive NSs created in binary NS mergers, finding that such stars collapse $\sim10\,{\rm s}$ -- $4.4\times10^{4}\,{\rm s}$ ($95\%$ confidence) after the merger. This directly impacts observations targeting NS remnants of binary NS mergers, providing the optimal window for high time resolution radio and $X$-ray follow-up of SGRBs and gravitational wave bursts.
	
\end{abstract}

\begin{keywords}
gamma-ray burst: general --- radio continuum: general --- stars: magnetars --- black hole physics --- equation of state
\end{keywords}

\section{Introduction}

Fast radio bursts \citep[FRBs; ][]{lorimer07,keane12,thornton13} are among the most exciting astronomical discoveries of the last decade. They are intense, millisecond-duration broadband bursts of radio waves so far detected in the 1.2\,GHz to 1.6\,GHz band, with dispersion measures between $300$\,cm$^{-3}$\,pc and $1200$\,cm$^{-3}$\,pc. FRBs are not associated with any known astrophysical object. Their anomalously large dispersion measures given their high galactic latitudes, combined with the observed effects of scattering consistent with propagation through a large ionised volume \citep{lorimer07,thornton13}, suggest a cosmological origin for FRBs \citep[although, see][]{sarah11,loeb13,bannister14}. 

Numerous physical mechanisms have been suggested to produce FRBs at cosmological distances, including magnetar hyperflares \citep[e.g.,][]{popov13}, binary white dwarf \citep{kashiyama13} or neutron star \citep[NS;][]{totani13} mergers, and the collapse of supramassive NSs to form black holes \citep{falcke13}. Here, we focus on the latter mechanism, termed the `blitzar' model.  A supramassive NS is one that has a mass greater than the non-rotating maximum mass, but is supported from collapse by rotation. As these stars spin down, they lose centrifugal support and eventually collapse to black holes. When magnetic field lines cross the newly-formed horizon, they snap violently, and the resulting outwardly-propagating magnetic shock dissipates as a short, intense radio burst \citep{falcke13,dap13}. 

The merger of two NSs is one possible formation channel for supramassive NSs. In general, there are three possible outcomes of such a merger, which, for a given NS equation of state (EOS), depend on the nascent NS mass, $M_p$, and angular momentum distribution. These outcomes are as follows:
\begin{enumerate}
	\item If $M_{p}\leq M_{\rm TOV}$, where $M_{\rm TOV}$ is the maximum non-rotating mass, the NS will settle to an equilibrium state that is uniformly rotating and eternally stable \citep[e.g.,][]{giacomazzo13}.
	\item If $M_{p}>M_{\rm TOV}$, and if magnetic braking has caused the star to rotate uniformly (see below), it will survive for $\gg1$\,s as a supramassive star until centrifugal support is reduced to the point where the star collapses to a black hole \citep[e.g.,][]{duez06}.
	\item If $M_{p}$ is greater than the maximum mass that may be supported by uniform rotation, the NS may either instantly collapse to a black hole or survive for $10-100$\,ms as a hypermassive star supported by differential rotation and thermal pressures \citep[e.g.,][]{baiotti08,kiuchi09b,rezzolla11,hotokezaka13}. 
\end{enumerate}  

Gamma-ray bursts (GRBs) with short durations ($\lesssim2$\,s) and hard spectra are associated with the coalescences of compact object binaries, i.e., either NS-NS or NS-black hole mergers \citep{nakar07,lee07,berger13}. However, the emission mechanisms of short GRBs (SGRBs) are not well understood. From a theoretical standpoint, a short-lived, collimated, relativistic jet can be launched from both black hole \citep[e.g.,][]{rezzolla11} and NS \citep[e.g.,][]{metzger08a} remnants of compact binary coalescences, as necessitated by the `relativistic fireball' model of prompt GRB emission \citep[e.g.,][]{piran99,nakar07,lee07}. Jet launching through magnetic acceleration requires a short-lived ($\sim0.1-1$\,s) accretion disk, and toroidal and small-opening-angle poloidal magnetic fields that are both $\sim10^{15}$\,G \citep[e.g.,][]{komissarov09}. 
Numerical simulations of the merger of two 1.5$M_{\odot}$ NSs with $10^{12}$\,G poloidal magnetic fields \citep{rezzolla11} result in a black hole remnant with the necessary conditions, where the magnetic field is amplified through magnetohydrodynamic instabilities, to launch a jet with an energy output of $\sim1.2\times10^{51}$\,erg. This is consistent with SGRB observations \citep{nakar07,lee07}.
Simulations of a lower-mass NS binary coalescence by \citet{giacomazzo13} resulted in a stable millisecond `protomagnetar', although the small-scale instabilities required to amplify the magnetic field to $\sim10^{15}$\,G were unresolved. On the other hand, high-resolution simulations of isolated NSs \citep[e.g.,][]{duez06} show that magnetic field amplification to $\sim10^{15}$\,G and accretion disk formation is possible for nascent protomagnetars, suggesting that protomagnetars can power prompt SGRB emission through magnetically accelerated jets. Baryon-free energy deposition through neutrino-antineutrino annihilation, driven by accretion onto protomagnetars, is another mechanism to power prompt SGRBs \citep{metzger08a}. In this paper, we assume that nascent protomagnetars formed through binary NS coalescences can power prompt SGRB emission, although we note that a deeper understanding is still required.

The most attractive characteristic of the protomagnetar model for SGRBs is its ability to explain features of the $X$-ray afterglows. Lasting a few hundred seconds following the initial burst, these features are either extremely bright \citep[up to 30 times the fluence of the prompt emission;][]{perley09} and variable on timescales comparable to prompt emission variability, or smoothly decaying plateaus \citep{rowlinson13}. Both features are difficult to explain through, for example, fall-back accretion onto black hole central engines \citep[][and references therein]{metzger08a,bucciantini12}, although see \citet{siegel2014}. However, the bright $X$-ray afterglows are explained by the effects of surrounding material on outflows from millisecond protomagnetars with strong magnetic fields \citep{metzger08a,bucciantini12}. The plateau phases observed in 65\% of SGRB $X$-ray afterglow lightcurves are consistent with electromagnetic spin-down energy losses from protomagnetars with dipolar magnetic field strengths of $B_{p}>10^{15}$\,G and rotation periods at the beginning of the X-ray emitting phase of $p_{0}\sim1$\,ms \citep{zhang01,rowlinson13}. 
Of this population, 39\% show an abrupt decline in $X$-ray flux within 50$-$1000\,s of the SGRB event, which is interpreted by \citet{rowlinson13} as the gravitational collapse of supramassive protomagnetars to black holes. A similar interpretation is given to abrupt declines in the $X$-ray afterglows of long GRBs \citep{troja07,lyons10}.

In this paper, motivated by the possible connection between the protomagnetar model for SGRBs and the blitzar scenario for FRBs \citep[e.g.,][]{zhang14}, we present a robust estimate of the possible lifetimes of supramassive protomagnetars created in binary NS mergers.\footnote{We stress that FRBs associated with SGRBs are likely to represent only a subset of the FRB population \citep{zhang14}.} The observation of an FRB associated with a sharp decline in the plateau phase of an SGRB $X$-ray lightcurve would provide powerful evidence for the protomagnetar model for SGRBs, as well as the blitzar model of FRBs. FRB/SGRB associations may also be used to characterise the intergalactic baryon content of the Universe \citep{deng14} through measurements of FRB dispersion measures and SGRB redshifts. Our results also provide a quantitative guide to interpreting observations of declines in SGRB $X$-ray lightcurves \citep{rowlinson13} in the context of supramassive protomagnetar collapse. In particular, secure measurements of the lifetimes of supramassive NSs produced in binary NS mergers allow for interesting new constraints to be placed on the equation of state (EOS) of nuclear matter \citep{lasky13c}.

Mergers of binary NSs are prime candidate sources for ground-based gravitational wave (GW) interferometers \citep{abadie10}.  However, the collimated nature of SGRBs \citep[e.g.,][]{burrows06} implies not all GW events will be associated with SGRBs.  Some alternative electromagnetic signals of GW bursts \citep{zhang13b,gao13} rely on the births of stable or supramassive NSs acting as central engines, implying the timescales for these signals are sensitively related to the lifetimes of the nascent NSs and therefore the calculations performed herein.

We show that, if a binary NS merger remnant does not collapse in the first $\sim4.4\times10^{4}\,{\rm s}$ after formation, it is unlikely to collapse at all (97.5\% confidence).  That is, almost all supramassive NSs born from binary NS mergers will collapse to form black holes within $\sim4.4\times10^{4}\,{\rm s}$. In \S2, we summarise our method for calculating the range of collapse times of merger remnants. We check the consistency of this method against a sample of SGRBs with $X$-ray plateaus from \citet{rowlinson13} in \S3. In \S4, we generalise our calculations to the full population of binary NS merger remnants, and we present our conclusions in \S5.

\section{Predicting supramassive neutron star collapse times}

The problem of identifying the spin-down torques on nascent NSs has been extensively studied over the last 50 years, with the primary aim of understanding any possible gravitational wave emission \citep[for a review, see][]{andersson03}. Two spin-down mechanisms are generally considered: electromagnetic (EM) radiation driven by the rotating NS magnetic field, and GW emission driven by bar-mode instabilities, stellar oscillation modes or magnetic field deformations.  Bar-modes in a rapidly rotating, gravitationally bound fluid body may be excited at large values of $T/|W|$, where $T$ is the rotational kinetic energy and $W$ is the gravitational potential energy. Non-axisymmetric ellipsoids are the only equilibrium configurations for $T/|W|\gtrsim0.27$ \citep{chandrasekhar69}; this limit is reduced to $T/|W|\gtrsim0.255$ in full General Relativity calculations \citep{baiotti07}.
For lower values of $T/|W|$, secular instabilities driven either by viscosity \citep{roberts63} or gravitational radiation reaction \citep[the `CFS' instability,][]{chandrasekhar70b,friedman75} can also lead to the formation of a bar mode. In particular, secular bar modes are excited for $T/|W|\gtrsim 0.14$, where $T/|W|\approx0.14$ is the bifurcation point between equilibrium sequences of Maclaurin spheroids and the non-axisymmetric Jacobi and Dedekind ellipsoids \citep[see][]{andersson03}.
The high temperatures of nascent NSs imply low viscosities such that secular bar modes are likely excited through the CFS instability \citep{lai95}.


\begin{figure*}
\centering
\includegraphics[angle=-90,scale=0.7]{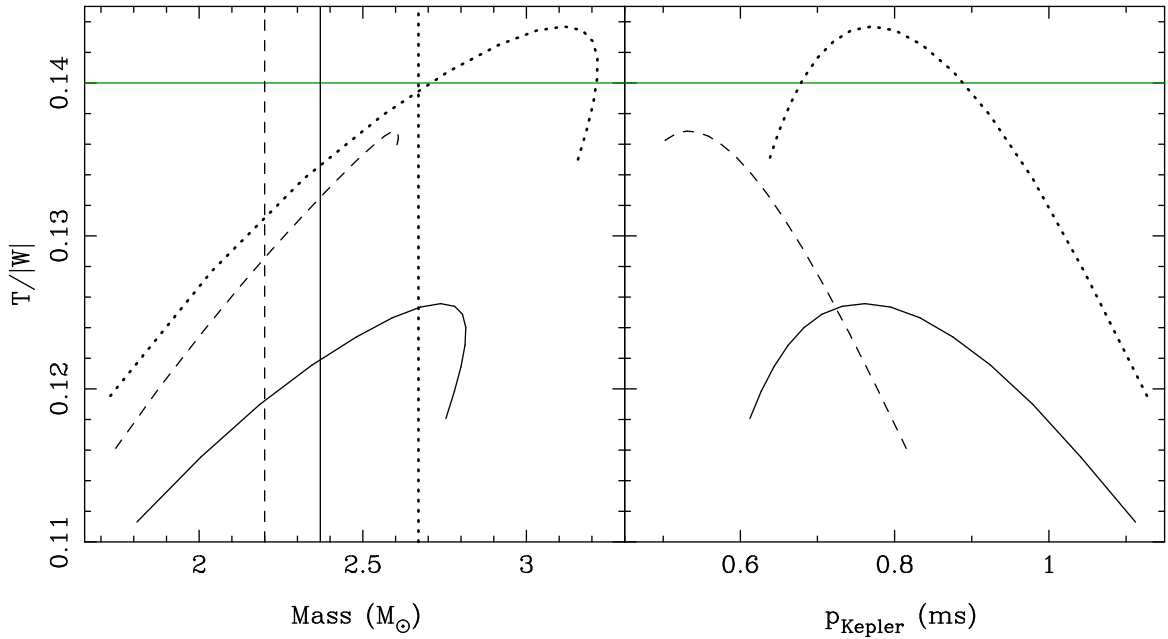}
\caption{Equilibrium sequences of $T/|W|$ at Keplerian break-up rotation velocity for three EOSs: GM1 (solid curves), APR (dashed curves) and AB-N (dotted curves). We plot $T/|W|$ against the NS mass, $M_{p}$, in the left panel, and against $p_{\rm Kepler}$ in the right panel. The vertical lines in the left panel indicate the values of the maximum non-rotating mass, $M_{\rm TOV}$ for each EOS, using the same line styles as the curves. The horizontal green line in both panels at $T/|W|=0.14$ indicates the approximate value of $T/|W|$ above which NSs are subject to secular instabilities.}
\end{figure*}

To date, the evolution of supramassive binary NS merger remnants has not been calculated in numerical relativity simulations. \citet{giacomazzo13} showed that a low-mass binary NS merger, with remnant mass $M_{p}=2.36M_{\odot}$, resulted in a stable, differentially rotating, bar-deformed NS within 55\,ms. In general, however, if a nascent supramassive NS exhibits a dynamical bar mode, the resulting gravitational radiation will cause the mode to be suppressed on a timescale of $\lesssim10$\,s \citep[e.g.,][]{baiotti07,baiotti08,hotokezaka13}. Differential rotation within the equilibrium configuration will be suppressed on the Alfv\'en timescale \citep{baumgarte00,shapiro00}, which is approximated as
\begin{equation}
t_{\rm Alfven} \approx 0.15\left(\frac{B_{p}}{10^{15}\,{\rm G}}\right)\left(\frac{M_p}{M_{\odot}}\right)^{1/2}\left(\frac{R}{10\,{\rm km}}\right)^{-1/2}\,{\rm s}. 
\end{equation}
For the protomagnetar model of SGRBs under consideration here, $B_{p}\sim10^{15}$\,G, which suggests that differential rotation is not expected to be significant over timescales $\gg1$\,s. We note, however, that this expression for the Alfv\'en timescale is likely to be a lower limit given that complex magnetic field structures are expected within protomagnetars \citep[e.g.,][]{giacomazzo13}. While dynamical bar-modes excited through shear instabilities may be present for $T/|W|<0.255$ \citep{corvino10}, this mechanism requires significant differential rotation and does not apply on timescales longer than the Alfv\'en timescale. 
Hence, for a protomagnetar to survive for $\gg10$\,s, as suggested by SGRB $X$-ray lightcurve observations \citep{bucciantini12,rowlinson13}, it must be truly supramassive, i.e., centrifugally supported against collapse to a black hole by uniform rotation. 

We further argue that such a supramassive protomagnetar is unlikely to be subject to the secular bar-mode instability. Throughout this work, we consider three representative EOSs with moderate $M_{\rm TOV}$ that are consistent with current galactic NS observations and the results of \citet{lasky13c}: APR \citep[][$M_{\rm TOV}=2.20\,M_\odot$, $R=10.00\,{\rm km}$, where $R$ is the stellar radius]{APR}, GM1 \citep[][$2.37\,M_\odot$, $12.05\,{\rm km}$]{GM1}, and AB-N \citep[][$2.67\,M_\odot$, $12.90\,{\rm km}$]{arnett77}. For each EOS, we used the general relativistic NS equilibrium code {\tt RNS} \citep{stergioulas95} to generate equilibrium sequences of $T/|W|$ at the Keplerian break-up rotation period $p_{\rm Kepler}$, assuming uniform rotation. These sequences are plotted against $M_{p}$ and $p_{\rm Kepler}$ in the left and right panels respectively of Fig.~1. We indicate the values of $M_{\rm TOV}$ for each EOS in the left panel, and also indicate the value of the secular bifurcation point $T/|W|\approx0.14$ as a horizontal green line in both panels. Only for EOS AB-N are values of $T/|W|>0.14$ possible, in particular for $M_{p}\gtrsim M_{\rm TOV}$. Hence, assuming EOS AB-N, supramassive protomagnetars may develop a secular bar instability. However, the growth timescale of the CFS-driven bar mode instability can be estimated from \citet[][their Eq.~2.7]{lai95}.  We find a minimum growth time of $\gtrsim10^{6}$\,s for EOS AB-N which, as we show below, is significantly longer than the maximum lifetime of a supramassive NS for this EOS given EM spin-down alone.


\citet{lasky13c} calculated the collapse time for a supramassive NS under the assumption of vacuum dipole spin down, showing that this depends on the EOS, the initial magnetic field strength, $B_p$, the initial spin period, $p_0$, and the supramassive NS mass, $M_p$. The collapse time is given by
\begin{equation}
t_{\rm col}=\frac{3c^3I}{4\pi^2B_p^2R^6}\left[\left(\frac{M_p-M_{\rm TOV}}{\alpha M_{\rm TOV}}\right)^{2/\beta}-p_0^2\right].\label{eq:tcol}
\end{equation}
Here, $M_{\rm TOV}$, $R$ and $I$ are the non-rotating maximum mass, radius and moment of inertia for a given EOS, and $\alpha$ and $\beta$ define the NS maximum mass as a function of spin period given by $M_{\max}(p)=M_{\rm TOV}\left(1+\alpha p^\beta\right)$. In Newtonian gravity, $\beta=-2$ and $\alpha$ is determined by the stellar properties. \citet{lasky13c} calculated general relativistic equilibrium sequences of $M_{\rm max}(p)$ using  {\tt RNS} \citep{stergioulas95} to evaluate $\alpha$ and $\beta$ for each EOS.  For EOS APR, $\alpha=3.03\times10^{-11}\,{\rm s}^{-\beta}$ and $\beta=-2.95$; EOS GM1 has $\alpha=1.58\times10^{-10}\,{\rm s}^{-\beta}$ and $\beta=-2.84$; and EOS AB-N has $\alpha=1.12\times10^{-11}\,{\rm s}^{-\beta}$ and $\beta=-3.22$.

As supramassive protomagnetars born in binary NS mergers are likely to neither be differentially rotating nor losing significant amounts of energy and angular momentum to GWs, we adopt this expression for the collapse time in the present work. We stress, however, that $B_{p}$ and $p_{0}$ for a protomagnetar are taken to be the magnetic dipole strength and spin period respectively at the epoch after which dynamical instabilities and differential rotation are suppressed. 

Our assumption of vacuum dipole spin down requires justification. The specific assumption of a dipolar magnetic field structure for binary NS merger remnants is unlikely to be the correct representation. The more pronounced decay with radial distance of high-order multipole components of the field implies that the dipole component will dominate on large spatial scales, and therefore possibly dominate the electromagnetic spin-down. Nonetheless, we emphasise that our calculations in this paper are entirely reliant on this assumption, noting that a magnetic field configuration partitioned into numerous multipoles will result in a slower spin-down rate than a purely dipolar configuration. 

Magnetic field induced deformities are not expected to cause significant gravitational wave emission over longer timescales \citep{haskell08}.  On the other hand, fallback accretion torques may decrease the spin-down \cite[e.g.,][]{piro11}, but the small mass of binary NS merger ejecta \citep{rezzolla11,hotokezaka13,giacomazzo13} combined with accretion disk heating effects \citep[][and references therein]{bucciantini12} imply that the accretion torques should be negligible.
 Finally, while Eq.~(2) includes the specific assumption of orthogonal rotation and magnetic axes, we note that $t_{\rm col}\propto \sin^{-2}(\theta)$, where $\theta$ is the angle between the magnetic and rotation axes \citep{shapiro83}.   This term introduces a factor of order unity in the collapse time, that we ignore for the remainder of this paper.

The collapse time is strongly dependent on the difference between $M_{p}$ and $M_{\max}(p)$ for a given EOS. Given a protomagnetar specified by $B_{p}$ and $p_{0}$, and an EOS which specifies $M_{\rm TOV}$, $R$, $I$, $\alpha$ and $\beta$, we use the observationally-derived distribution of NS masses in galactic binary NS systems to predict the distribution of $t_{\rm col}$. From \citet{kiziltan13}, $M_{\rm BNS}\sim N(\mu_{\rm BNS}=1.32\,M_{\odot},\sigma_{\rm BNS}=0.11)$, where $N(\mu_{\rm BNS},\sigma_{\rm BNS})$ specifies a normal distribution with mean $\mu_{\rm BNS}$ and standard deviation $\sigma_{\rm BNS}$. Binary NS mergers are believed to approximately conserve rest mass ($M_{\rm rest}$), with $\lesssim0.01M_\odot$ of material ejected during the merger \citep{rezzolla11,hotokezaka13,giacomazzo13}.  We use the approximate conversion $M_{\rm rest}=F_{\rm rest}(M_{g})=M_{g}+0.075M_{g}^{2}$ \citep{timmes96}, where $M_g$ is the gravitational mass, to derive $M_{p}$:
\begin{equation}
M_{p}=F_{\rm rest}^{-1}[F_{\rm rest}(M_{\rm BNS,\,1})+F_{\rm rest}(M_{\rm BNS,\,2})].
\end{equation}
Here, $M_{\rm BNS,\,1,\,2}$ are random variables from the distribution $N(\mu_{\rm BNS},\sigma_{\rm BNS})$. By generating a large sample of protomagnetar masses, we produce a large sample of $t_{\rm col}$ trials, allowing us to calculate the cumulative distribution function (CDF), $\hat{f}(t_{\rm col})$, for the collapse time of a particular protomagnetar assuming a particular EOS.

\section{Candidate protomagnetar collapse times} 

\citet{rowlinson13} analysed observations of plateaus in the $X$-ray lightcurves of SGRBs, fitting protomagnetar spin-down emission models \citep{zhang01} to the data to infer $B_{p}$ and $p_{0}$. We compare our predictions for the range of possible $t_{\rm col}$ with these observations. We restrict our attention to the sample of eight SGRBs from \citet{rowlinson13} that have reliable redshift measurements and show strong evidence for NS central engines [see Table~1 of \citet{lasky13c}].\footnote{There is debate as to the redshift of GRB\,070809.  \citep{rowlinson13} fit the NS spin-down model assuming $z=0.219$ \citep{perley08}. While a lower probability host galaxy has been identified at $z=0.473$ \citep{berger10}, we continue to use the \citet{rowlinson13} fit in this work, noting that the inferred values of $B_p$ and $p_0$ change for different redshift identifications.} Four of these show abrupt declines in their X-ray lightcurves between 90\,s and 330\,s after the initial GRB, and the remaining four show no departure from the plateau phase for $5\times10^{4}-5\times10^{5}$\,s, after which the lightcurve is no longer distinguishable from measurement noise.

\begin{figure}
\centering
\includegraphics[scale=0.39,angle=-90]{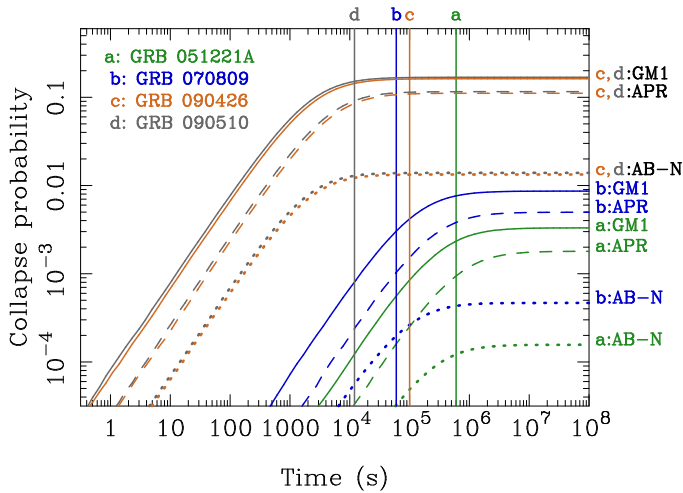}
\caption{Predicted probability of collapse as a function of time for four protomagnetars in the \citet{rowlinson13} sample that are not observed to collapse. The curves indicate the probabilities of $0<t_{\rm col}<t$ for a time $t$ after the initial NS-NS mergers. We calculate the probabilities for three EOSs: GM1 (solid curves), APR (dashed curves) and AB-N (dotted curves). The thin vertical lines indicate the (restframe) times of the last observations of the $X$-ray lightcurves, that in all cases are consistent with the continued existence of the protomagnetars.}
\end{figure}

The four NS candidates that were not observed to collapse will follow one of two evolutionary paths:  if $M_{p}>M_{\rm TOV}$, the NS will eventually collapse to form a black hole or, if $M_{p}\le M_{\rm TOV}$, the NS will be eternally stable. We calculate CDFs, $\hat{f}(t_{\rm col})$, for each of these objects using the estimated values of $B_{p}$ and $p_{0}$ from \citet{rowlinson13}. We reject cases for which $M_{p}>M_{\max}(p_{0})$ as these either instantly collapse to form a black hole or form a dynamically unstable hypermassive NS. We keep track of the number of cases not included in the samples for which $M_{p}\leq M_{\rm TOV}$, in order to appropriately normalise $\hat{f}(t_{\rm col})$. 

In Fig.~2, we plot $\hat{f}(t_{\rm col})$ for each of the four SGRBs that are not observed to collapse.  For each SGRB we consider the three representative EOSs introduced in \S2. We do not consider the experimental errors in the determination of $B_{p}$ and $p_{0}$ as they do not significantly affect our results \citep{lasky13c}.  The vertical lines in Fig.~2 are the restframe times of the last observation of the $X$-ray lightcurves.

The curves in Fig.~2 indicate the probabilities of collapse before a time $t$ after the bursts, i.e., the probabilities of $0<t_{\rm col}<t$. Consider GRBs 090426 and 090510 in Fig.~2: if these candidate NSs were born with $M_p>M_{\rm TOV}$ for any EOS, collapse to a black hole {\it should} have been observed while the $X$-ray afterglows were being monitored.  Fig.~2 also shows that the probability that these two candidate NSs were born with $M_p\le M_{\rm TOV}$ is between 90 -- 99\% (depending on the EOS) implying stable NSs probably exist in the merger remnants.  On the other hand, if the candidate NSs born in GRBs 051221A and 070809 had $M_p>M_{\rm TOV}$, there is a $\sim$30--60\% chance (depending on the EOS and the GRB properties, indicated by the intersections between the relevant curves and vertical lines indicating observation times) that collapse occurred after the observations were concluded rather than during the observations. However, these NSs have a $>99\%$ chance of having been born with $M_p\le M_{\rm TOV}$.

The likelihoods of these four candidate protomagnetars being supramassive, as opposed to eternally stable, are low for all EOSs, which is in agreement with the lack of observational evidence for collapse. It is instructive to consider these likelihoods for the four candidate protomagnetars of \citet{rowlinson13}, discussed above and in \citep{lasky13c}, that exhibited sharp declines in their $X$-ray lightcurves indicative of collapse. For EOS GM1, which provides the greatest chance of collapse [see also \citet{lasky13c}], we find a probability for $0<t_{\rm col}<\infty$ of 0.2\% for the GRB\,080905A, 15\% for GRB\,060801, 18\% for GRB\,070724A and 60\% for GRB\,101219A.  The latter three predictions are broadly consistent with the \citet{rowlinson13} observations.

GRB\,080905A is a strange case that has been discussed extensively. \citet{fan13} interpreted the slow spin-period \citep{rowlinson13} as evidence that gravitational radiation contributed significantly to the spin-down. However, as discussed in \S2, it is difficult to identify a mechanism to generate gravitational radiation on these timescales.
It is possible that other issues, such as sub-optimal efficiency of turning rotational energy into electromagnetic energy \citep{rowlinson13} and the inferred redshift \citep{gruber12,howell13} are also critical.  Each of these effects change the inferred values of $p_0$ and $B_p$, and hence the probability of collapse.

\section{Collapse time distributions}

We now predict the range of possible collapse times, $t_{\rm col}$, for supramassive stars created in binary NS mergers, using the techniques discussed above for calculating $t_{\rm col}$ for a specific NS and EOS. This requires knowledge of the distributions of $p_{0}$ and $B_{p}$ expected for binary NS merger remnants. 

Local, high-resolution numerical simulations indicate that small-scale turbulent dynamos are expected to act throughout a binary NS merger, amplifying internal magnetic fields to $\sim$10$^{16}$\,G on timescales shorter than the merger timescale \citep{obergaulinger10,zrake13}.  Global simulations of binary NS mergers produce similar field strengths through magnetorotational instabilities \citep[e.g.,][]{price06,duez06,siegel13}, which are consistent with the inferred field strengths from the observations of \citet{rowlinson13}.  We assume all protomagnetars have poloidal magnetic field strengths between $10^{15}$\,G and $3\times10^{16}$\,G, where the lower limit is inferred from the need for collimated jets to explain the burst emission \citep{thompson07}, and the upper limit is based on magnetic field stability arguments \citep{metzger11}. 

A supramassive star of mass $M_{p}$ can have $p_{0}$ between the minimum (Keplerian) spin period, $p_{\rm Kepler}$, and the maximum spin period, $p_{\rm max}$, that permits the star to remain metastable to collapse. 
From Eq.~(\ref{eq:tcol}) the maximum spin period is given by 
\begin{equation}
p_{\rm max}=\left(\frac{M_{p}-M_{\rm TOV}}{\alpha M_{\rm TOV}}\right)^{1/\beta}.
\end{equation}

We generate random samples of supramassive protomagnetar masses, $M_{p}$, for each of the three EOSs we consider, i.e., with $M_{\rm TOV}<M_{p}<M_{\rm max}$. For each mass value, we draw values of $p_{0}$ and $B_{p}$ from log-uniform distributions bounded by $p_{\rm Kepler}\le p_0\le p_{\rm max}$ and $1\le B_{p}/10^{15}\,{\rm G}\le30$. 
Values of $p_{\rm Kepler}$ for each mass value are obtained by cubic interpolation of the equilibrium Keplerian sequences shown in the left panel of Fig.~1, and values of $p_{\rm max}$ are obtained using Eq.~(4).  
We discuss the effects of choosing these distributions below.  Having calculated the CDF, $\hat{f}(t_{\rm col})$, for each sample, we show the estimates of the predictive probability density functions (PDFs) for $t_{\rm col}$ in Fig.~3 for each EOS.

\begin{figure}
\centering
\includegraphics[scale=0.34,angle=-90]{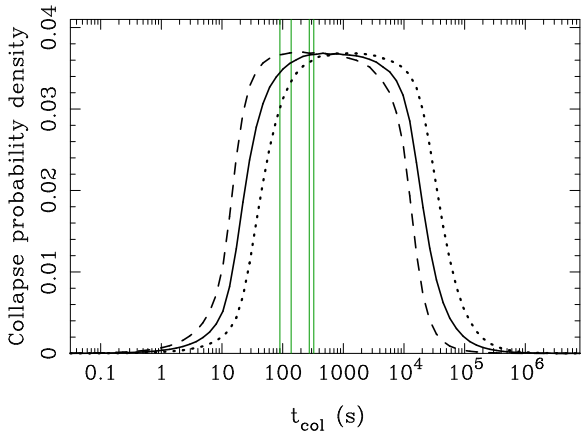}
\caption{PDFs for $t_{\rm col}$ derived from $\hat{f}(t_{\rm col})$ for three possible EOSs: GM1 (solid curves), APR (dashed curves) and AB-N (dotted curves). We also show (vertical green lines) the observed values of $t_{\rm col}$ for four SGRB remnants with secure redshift measurements observed to collapse by \citet{rowlinson13}.}
\end{figure}

\begin{figure}
\centering
\includegraphics[scale=0.34,angle=-90]{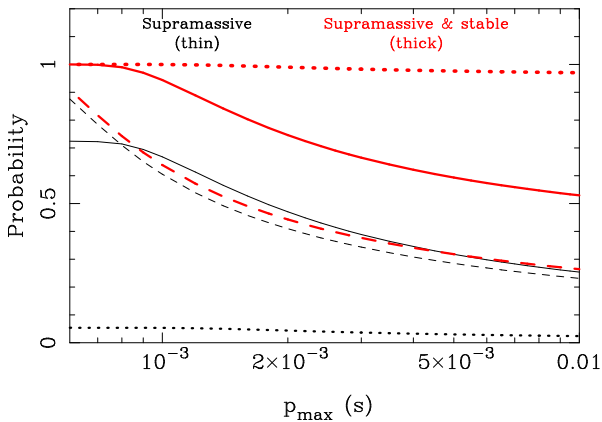}
\caption{The fraction of binary NS mergers that result in supramassive NSs [i.e., with $M_{\rm TOV}<M_p<M_{\rm max}(p_{0})$; thin black curves] and supramassive together with eternally stable NSs [$M_p<M_{\rm max}(p_{0})$; thick red curves], as functions of the maximum protomagnetar spin period. The linestyles correspond to the different EOSs: GM1 (solid curves), APR (dashed curves) and AB-N (dotted curves).}
\end{figure}

Fig.~3 shows that the expected collapse times for supramassive protomagnetars formed in binary NS mergers are relatively short. Applying equal weighting to each EOS, we find that $t_{\rm col}$ is in the range $\sim10\,{\rm s}$ -- $4.4\times10^{4}\,{\rm s}$ with 95\% confidence. The observed restframe collapse times from \citet{rowlinson13}, shown in Fig.~3 as vertical green lines, are consistent with all EOSs, although they appear somewhat smaller than what may generally be expected. 

The PDFs in Fig.~3 encompass a wide range of uncertainties in the collapse times. While we show the effects of the choice of EOS, the PDFs are also affected by uncertainties in $M_{p}$, $p_{0}$ and $B_{p}$. The lower limits of the curves are set by the maximum allowed value of $B_{p}$, while the forms of the intrinsic distributions of $p_{0}$ and $B_{p}$, for which we have assumed the least constraining possibilities, affect the shapes of the curves. Significantly more precise predictions of $t_{\rm col}$ are possible for individual candidate protomagnetars with inferred values of $p_{0}$ and $B_{p}$. As pointed out in \citet{lasky13c}, however, the prediction then depends sensitively on the assumed EOS.

We have so far only considered binary NS mergers that result in supramassive stars, i.e., that have $M_{\rm TOV}<M_p<M_{\rm max}(p_0)$ and $0<t_{\rm col}<\infty$. Of the entire population of binary NS mergers, we find that the probabilities for the newly-created protomagnetars to fall into this category are 5\% for EOS AB-N, 72\% for EOS GM1 and 97\% for EOS APR. Given the sensitivity of these probabilities to the EOS, we {\it cannot} independently predict the rate of production of supramassive NSs from binary NS mergers.

It is useful to consider how the probabilities of supramassive NS production vary as functions of a universal $p_{\rm max}$. We note that these probabilities do not depend on the $B_{p}$-distribution.  In Fig.~4 we repeat our calculations for fixed values of $p_{\rm max}$, rather than allowing $p_{\rm max}$ to vary depending on $M_{p}$ for each trial. For each EOS we show two curves: the probability of a binary NS merger resulting in a supramassive NS [i.e., with $M_{\rm TOV}<M_p<M_{\rm max}(p_{0})$; thin black curves] and the probability of a merger resulting in either a supramassive or an eternally stable NS [i.e., with $M_p<M_{\rm max}(p_{0})$; thick red curves]. We compare the results in Fig.~4 with the observed fractions of SGRBs with $X$-ray plateaus, and those that also show abrupt declines in $X$-ray flux \citep{rowlinson13}, both of which we consider as upper limits on the true fractions. It is unlikely that EOS AB-N is consistent with the $\sim 40-65$\% of SGRBs observed to have $X$-ray plateaus \citep{rowlinson13}. It is also unlikely that EOS APR is consistent with the $\sim 39$\% of SGRBs with $X$-ray plateaus that exhibit abrupt declines. However, EOS GM1 is consistent with both fractions, in particular for $p_{\rm max}>8$\,ms, a result that is also in accordance with the \citet{usov92} suggestion that $p_{\rm max}\sim10$\,ms.

\section{Discussion and conclusions}

We predict the collapse times of supramassive protomagnetars born from the merger of two neutron stars (NSs).  We show that, if the protomagnetar has not collapsed within $4.4\times10^{4}\,{\rm s}$, the probability that it will ever collapse is small; quantitatively, $P(t_{\rm col}>4.4\times10^{4}\,{\rm s})=0.025$. We also consider the dependence on the assumed NS equation of state (EOS) and the distribution of initial spin periods of the fractions of binary NS mergers that result in supramassive and eternally stable NSs. Of the scenarios we consider, only EOSs similar to GM1 and $p_{\rm max}>8$\,ms are consistent with the current short gamma-ray burst (SGRB) $X$-ray lightcurve sample \citep{rowlinson13}. 

Our results strongly impact observations targeting protomagnetars created in binary NS mergers. We consider these observations in turn:
\begin{enumerate}
\item \textit{Fast radio bursts (FRBs) from blitzars associated with SGRBs.} \citet{falcke13} posit that the collapse of a supramassive NS causes the emission of an intense radio burst. These authors further suggest that $\sim10^{3}$\,yr may be required to sufficiently clear the NS environments to allow the efficient outward propagation of blitzar radio emission. However, \citet{zhang14} found that GRB blast waves and the shocked circum-burst media are likely to have plasma oscillation frequencies significantly lower than FRB emission frequencies, indicating that blitzars occurring shortly after SGRBs will be visible. Our results show that radio follow-up observations of SGRBs to detect FRBs have an optimal detection window between 10\,s and $4.4\times10^{4}$\,s after the initial burst. This is consistent with the calculation of \citet{zhang14}, who suggested radio follow-up observations commencing 100\,s following the initial burst. We note that from the SGRB sample of \citet{rowlinson13}, it is likely that between 15\% and 25\% of SGRBs will result in supramassive stars. 
\item \textit{$X$-ray SGRB follow-up.} The theoretical framework that we use to calculate the collapse time window provides a guide to interpreting $X$-ray observations of SGRBs designed to be sensitive to abrupt declines in plateau phases of the lightcurves. The observation of an abrupt decline greater than $4.4\times10^{4}$\,s following the initial burst would be inconsistent with this framework. 
\item \textit{Electromagnetic counterparts to gravitational wave sources.} A large fraction of any non-SGRB signatures of binary NS mergers that rely on a protomagnetar central engine \citep{zhang13b,gao13} may have timescales associated with the distribution of supramassive NS collapse times (Fig.~3). Any such signatures with corresponding timescales, conversely, may be suggestive of binary NS mergers. 
\end{enumerate}

Our calculations have various uncertainties that require further study. The basic electromagnetic spin down formula we apply to calculate $t_{\rm col}$ in Eq.~(2) requires revision to fully model the spin down torques of binary NS merger remnants \citep{lasky13c}. Additional spin-down mechanisms cause $t_{\rm col}$ to reduce in general, and may affect the overall fractions of supramassive NSs created in binary NS mergers. We also do not account for changes in the gravitational masses, radii and moments of inertia of NSs as they spin down, although these effects are not likely to significantly change our results \citep[e.g., see][]{falcke13}. Our assumption of a simple dipole field structure orthogonally oriented to the rotation axis is also unlikely to be fully correct, although we do not account for this uncertainty in our predictions.  The biggest uncertainties included in our predictions of $t_{\rm col}$ come from the assumed distributions for $p_0$ and $B_p$ and the assumed EOS. We have, however, chosen the broadest possible priors on $p_0$ and $B_p$, and consider a variety of EOSs, implying the range of possible $t_{\rm col}$ values we find is conservative.

In conclusion, we strongly urge high time resolution radio and $X$-ray follow-up observations of short gamma-ray bursts, and in the future gravitational wave bursts, from binary neutron star mergers. The optimal window for these observations is between 10\,s and $4.4\times10^{4}$\,s after the initial bursts. These observations have the potential to provide strong evidence for the protomagnetar model of short gamma-ray bursts and the blitzar model of fast radio bursts.

\section*{Acknowledgements}
We thank the referee for extremely helpful comments that helped strengthen the paper. We further thank B. Haskell, E. Howell, L. Rezzolla, A. Rowlinson, W.-F. Fong and A. Melatos for useful discussions. VR is a recipient of a John Stocker Postgraduate Scholarship from the Science and Industry Endowment Fund. PDL is supported by the Australian Research Council Discovery Project (DP110103347).

\bibliographystyle{mn2e}
\bibliography{Bib}

\label{lastpage}

\end{document}